\documentclass[14pt,a4paper,twoside]{report}

\usepackage{fontspec}
\usepackage{babel}
\babelprovide[import, main]{russian}
\babelprovide[import]{english}

\babelfont{rm}[
    Extension      = .otf,
    UprightFont    = cmunrm,
    BoldFont       = cmunbx,
    ItalicFont     = cmunti,
    BoldItalicFont = cmunbi
]{CMU Serif}
\babelfont[russian]{rm}[
    Extension      = .otf,
    UprightFont    = cmunrm,
    BoldFont       = cmunbx,
    ItalicFont     = cmunti,
    BoldItalicFont = cmunbi
]{CMU Serif}
\babelfont[english]{rm}[
    Extension      = .otf,
    UprightFont    = cmunrm,
    BoldFont       = cmunbx,
    ItalicFont     = cmunti,
    BoldItalicFont = cmunbi
]{CMU Serif}

\usepackage{amsmath}
\usepackage{amssymb}
\usepackage{longtable}
\usepackage{indentfirst}
\usepackage[usenames]{color}
\usepackage{graphicx,array,amsfonts}
\usepackage{epstopdf}
\usepackage{wrapfig}
\usepackage{tabularx}
\usepackage{booktabs}
\usepackage{geometry}
\usepackage{enumitem}
\usepackage{titlesec}
\usepackage{cite}

\graphicspath{{img/}}
\usepackage{vwcol}

\usepackage{hyperref}
\hypersetup{
    pdfstartview=FitH, 
    linkcolor=black, 
    urlcolor=black, 
    colorlinks=true,
    pdfauthor={Anonymous}, 
    pdftitle={Approximation Algorithms for Capacitated Vehicle Routing Problems: A Comprehensive Survey}
}

\newcommand{\doi}{\texttt{DOI 10.52575/2687-0959-20XX-XX-X-X-XX}}
\newcommand{\magazinenumben}{Applied Mathematics \& Physics, XX(X): }

\newcommand{\god}{20XX. }
\newcommand{\pages}{XX--XX. }

\titleformat{\section}{\large\bfseries}{\thesection}{1em}{}
\titleformat{\subsection}{\normalsize\bfseries}{\thesubsection}{1em}{}

\begin{document}
	
{\small
	
\begin{vwcol}[widths={0.45,0.45},
	sep=.9cm, justify=flush,rule=0pt,indent=0.1em]
	УДК 519.8
	
	MSC 90C27, 68W25
	
	\doi
\end{vwcol}

}

\vskip - 0.25cm
\noindent \rule{\textwidth}{0.5pt}
\vskip - 0.5cm
\noindent\rule{\textwidth}{0.5pt}
\vskip 0.5 cm

{\small

\noindent Review Article

}

\begin{center}
	\textbf{\large{Approximation Algorithms for Capacitated Vehicle Routing Problems: A Comprehensive Survey}}
	\medskip

{\small

	\textbf{Yongyu Chen}\textsuperscript{1 \href{https://orcid.org/0000-0003-2756-380X}{\texttt{[ORCID]}}}
	\vskip 0.2cm
	%
	%
	\textsuperscript{1}Ural Federal University,\\
	Lenin Ave, 51, Yekaterinburg, Sverdlovsk Oblast, Russia, 620075\\
	\textcolor{blue}{chen@urfu.ru}
	

}

\end{center}

{\small

\noindent \textbf{Abstract.} The Capacitated Vehicle Routing Problem (CVRP) is a core NP-hard problem in the field of combinatorial optimization. It aims to plan optimal routes for a fleet of vehicles with uniform capacity, serving a set of customers with specific demands from a single depot, while minimizing the total travel distance. Due to its extensive applications in logistics, distribution, and supply chain management, CVRP has attracted significant research attention. Theoretically, the problem has been proven to be APX-hard, and in general metric spaces, approximate solutions of arbitrary precision cannot be obtained unless P=NP. These inherent complexities highlight the importance of developing approximation algorithms—finding solutions with provable performance guarantees in polynomial time. This paper aims to provide a systematic and comprehensive survey of the research progress on CVRP approximation algorithms. The paper first strictly defines CVRP and its key variants, and elucidates the sources of its fundamental computational complexity. Subsequently, the article deeply analyzes the core algorithmic frameworks and technical schools of thought in this field, including: the Iterated Tour Partitioning (ITP) framework that laid the foundation of the field and its latest improvements; the evolution of approximation schemes (PTAS/QPTAS) for geometric spaces such as Euclidean space; and modern algorithms for structured graphs like trees, planar graphs, and graphs with bounded highway dimension, with a particular focus on techniques based on metric embedding and linear programming relaxation. Finally, this paper summarizes the current best approximation ratios for various problem settings and systematically outlines the core unresolved open problems in the field, pointing out directions for future research.
	
	\smallskip
	
	
	\smallskip
	
	\noindent \textbf{Acknowledgements:} The work is supported ... (Specify grant if any)
	
	\smallskip
	
	\noindent \textbf{For citation:} [Author]. \god Approximation Algorithms for Capacitated Vehicle Routing Problems: A Comprehensive Survey. \magazinenumben \pages \doi 
	
}

\vskip 0.3cm
\noindent\rule{\textwidth}{0.5pt}
\vskip 0.3cm


\section{Introduction}

\subsection{Problem Definition and Background}

The Capacitated Vehicle Routing Problem (CVRP) is a classic combinatorial optimization problem in operations research and computer science. Its formal definition is as follows: Given a complete graph $G=(V, E)$, where the vertex set $V$ consists of a depot (vertex 0) and a set of $n$ customers $V_c = \{1, ..., n\}$. A symmetric cost function $c: E \rightarrow \mathbb{R}_{\ge 0}$ is defined on the edges of the graph, satisfying the triangle inequality. Each customer $i \in V_c$ has a known demand $d_i$, and there exists an infinite fleet of homogeneous vehicles, each with a capacity limit of $Q$.

The goal of CVRP is to find a set of simple cycles (i.e., ``routes'' or ``tours''), where each cycle starts from the depot and eventually returns to the depot. These cycles collectively constitute a partition of the customer set $V_c$. The core constraint is that the total demand of all customers served by any single cycle must not exceed the vehicle capacity $Q$. The ultimate objective is to minimize the total cost of all cycles. In many practical applications, the optimization objective is often lexicographical: first minimizing the number of vehicles used, and secondly minimizing the total travel distance.

The theoretical and practical significance of CVRP is profound. Since Dantzig and Ramser first proposed the problem in their seminal paper ``The Truck Dispatching Problem'' in 1959 \cite{dantzig1959truck}, it has become a core model for simulating real-world logistics and distribution networks. From parcel delivery in e-commerce to urban waste collection, and to supply chain management of industrial goods, efficient CVRP solutions can bring significant cost savings, as transportation costs typically account for a large portion of the final cost of goods.

\subsection{Relationship with Fundamental Combinatorial Optimization Problems}

The complexity of CVRP stems from its deep connection with two more fundamental combinatorial optimization problems—the Traveling Salesman Problem (TSP) and the Bin Packing Problem (BPP). This dual nature is key to understanding the design of CVRP approximation algorithms.

First, CVRP is a direct generalization of TSP. TSP aims to find the shortest Hamiltonian cycle that visits all customers and returns to the starting point. If the vehicle capacity $Q$ in CVRP is large enough to serve all customers at once (or if the capacity constraint is removed), the problem degenerates into a standard TSP. This relationship implies that any CVRP algorithm must solve an inherent path planning or ``ordering'' problem. Therefore, many advanced CVRP approximation algorithms utilize efficient TSP approximation algorithms as their core subroutines (note also recent advances in Asymmetric TSP \cite{svensson2020constant}).

Second, CVRP embeds the structure of the Bin Packing Problem. The Bin Packing Problem requires packing a set of items with different sizes into the minimum number of bins with identical capacity. If we imagine all customers in CVRP are located at the same geographic location (i.e., the travel cost between them is zero), then CVRP simplifies to how to partition customer demands (items) into the minimum number of vehicles (bins), which is exactly the Bin Packing Problem. This relationship reveals another core challenge of CVRP: the ``partitioning'' or ``clustering'' of customers.

This dual nature of CVRP—containing challenges of both ordering (TSP) and partitioning (Bin Packing)—has directly spawned two major schools of heuristic algorithm design. The first is ``cluster-first, route-second,'' which first partitions customers into several groups where the total demand of each group satisfies the capacity limit, and then solves an independent TSP for each group. The second is ``route-first, cluster-second,'' which first ignores capacity limits to calculate a ``giant'' TSP tour visiting all customers, and then splits this tour into several sub-segments satisfying capacity constraints. As will be detailed in subsequent sections, the latter is the theoretical basis of the famous Iterated Tour Partitioning (ITP) algorithm, which has dominated the development of the CVRP approximation algorithm field for decades.

\subsection{Computational Complexity and Motivation for Approximation Algorithms}

The computational complexity of CVRP is extremely high. When the number of customers a vehicle can serve is at least 3 ($Q \ge 3$), the problem is NP-hard \cite{hassin2005complexity, zhao2022improved}. More importantly, due to its connection with the Bin Packing Problem, CVRP is APX-hard. This means that even in the simplest one-dimensional Euclidean space, no Polynomial Time Approximation Scheme (PTAS) exists unless P=NP. A PTAS is a class of algorithms that, for any given $\epsilon > 0$, can find a solution with a cost no more than $(1+\epsilon)$ times the optimal solution within polynomial time.

Furthermore, for CVRP in general metric spaces (i.e., spaces where distances only satisfy the triangle inequality), approximation is even more difficult. Arora et al. \cite{arora1992proof} proved in 1992 that Metric TSP does not have a PTAS (unless P=NP). Since CVRP is a generalization of TSP, this inapproximability result also applies to CVRP. These profound hardness results indicate that for CVRP in the general case, we cannot expect to find polynomial-time algorithms that are arbitrarily close to the optimal solution. Therefore, the focus of research shifts to designing approximation algorithms with an approximation ratio of a certain constant $\alpha$, which can provide a suboptimal solution with a quality guarantee in polynomial time.

\subsection{Structure of This Review}

This review aims to systematically summarize the research achievements and core technologies in the field of CVRP approximation algorithms. The structure is arranged as follows:

\begin{itemize}
    \item \textbf{Section 2} will rigorously define various important variants of CVRP and explore the theoretical roots of its hardness.
    \item \textbf{Section 3} will detail the Iterated Tour Partitioning (ITP) framework that laid the foundation for the field, including its classical analysis and major theoretical breakthroughs in recent years.
    \item \textbf{Section 4} will focus on approximation schemes in geometric spaces such as Euclidean space, tracing the evolution from Arora's TSP framework to QPTAS and PTAS for CVRP.
    \item \textbf{Section 5} will discuss algorithms designed for graphs with special combinatorial structures (such as trees, planar graphs), highlighting modern techniques like metric embedding and linear programming relaxation.
    \item \textbf{Section 6} will summarize the current state-of-the-art approximation results and look forward to the core open problems that urgently need to be solved in this field, providing references for future research.
\end{itemize}

\section{Problem Formalization, Variants, and Fundamental Hardness}

To deeply understand the evolution of CVRP approximation algorithms, one must first clearly define the different variants of the problem and understand the precise sources of its hardness. The difficulty of the problem depends not only on the basic NP-hard property but is also closely related to the demand type and the structure of the underlying metric space.

\subsection{Key Variants of CVRP}

The standard model of CVRP has derived multiple variants in theoretical research and practical applications. These variants differ mainly in demand characteristics, depot structure, and metric space. The following table (Table 1) formally defines these key variants.

\begin{longtable}{p{0.2\textwidth} p{0.25\textwidth} p{0.45\textwidth}}
\caption{Key Variants of CVRP and Their Formal Definitions} \label{tab:variants} \\
\toprule
\textbf{Variant Category} & \textbf{Variant Name} & \textbf{Formal Definition and Key Attributes} \\
\midrule
\endfirsthead
\caption[]{Key Variants of CVRP (continued)} \\
\toprule
\textbf{Variant Category} & \textbf{Variant Name} & \textbf{Formal Definition and Key Attributes} \\
\midrule
\endhead
\bottomrule
\endfoot
\bottomrule
\endlastfoot
\textbf{Demand Type} & \textbf{Unsplittable Demand} & The entire demand $d_i$ of each customer $i$ must be fully satisfied by a single vehicle in a single tour. This is the standard and most challenging assumption of CVRP. \\
\cmidrule{2-3}
 & \textbf{Splittable Demand} & Allows the demand $d_i$ of a customer $i$ to be satisfied jointly by multiple vehicles or across multiple tours. This assumption usually reduces the complexity of the problem, making it easier to approximate. \\
\cmidrule{2-3}
 & \textbf{Unit Demand} & All customers have exactly the same demand. Typically, the total demand is normalized so that each customer's demand is 1, where $Q$ is a positive integer representing the vehicle capacity (i.e., the maximum number of customers it can serve). \\
\midrule
\textbf{Depot Structure} & \textbf{Single-Depot} & All vehicles start from a common depot and eventually return to that depot. This is the standard setting for CVRP. \\
\cmidrule{2-3}
 & \textbf{Multi-Depot (MDVRP)} & Multiple depots exist; vehicles can start from and return to different depots. This increases the complexity of assigning customers to vehicles (and depots). \\
\midrule
\textbf{Metric Space} & \textbf{General Metric} & Distances (costs) between customers and the depot only need to satisfy the three metric axioms: non-negativity, symmetry, and triangle inequality. This is the most universal setting. \\
\cmidrule{2-3}
 & \textbf{Euclidean Space} & Customers and the depot are points in a $d$-dimensional Euclidean space $\mathbb{R}^d$, with distance defined by the $L_2$ norm. This geometric structure allows for powerful techniques like recursive space decomposition. \\
\cmidrule{2-3}
 & \textbf{Tree Metric} & Customers and the depot are nodes of a tree graph, and distance is defined as the length of the shortest path between nodes in the graph. This highly structured space often allows for algorithms with better performance, even PTAS. \\
\cmidrule{2-3}
 & \textbf{Structured Graphs} & Includes Planar Graphs, Minor-Free Graphs, graphs with Bounded Treewidth, Bounded Highway Dimension, or Bounded Doubling Dimension. These graph classes aim to capture the sparsity or hierarchical structure of real-world networks (like road networks) and represent the frontier of current research. \\
\midrule
\textbf{Other Constraints} & \textbf{With Time Windows (CVRPTW)} & Service for each customer must begin within a specified time window. This introduces scheduling and time feasibility constraints, greatly increasing the problem difficulty. \\
\cmidrule{2-3}
 & \textbf{Pickup and Delivery (VRPPD/VRPB)} & Tours include not only ``delivery'' tasks from depot to customers but may also include tasks to ``pick up'' goods from customers and return to the depot (Backhauling, VRPB) or transfer goods between customers (Pickup and Delivery, VRPPD). \\
\end{longtable}

The combination of these variants defines a vast problem space. For example, a problem could be ``CVRP with unsplittable unit demands in Euclidean space'' or ``CVRPTW with splittable demands in a general metric space.'' The performance and applicability of algorithms are often closely related to these specific settings. When classifying problem difficulty, a general rule can be observed: the complexity of the problem is a multidimensional function determined jointly by the structure of the metric space and the demand type. Generally, the stronger the structural property of the metric space (e.g., from general metric space to bounded dimension graphs, to Euclidean space, and finally to trees), the easier the problem is to handle. Simultaneously, the looser the demand constraints (e.g., from unsplittable demand to unit demand, to splittable demand), the easier the problem is to approximate. Progress in this field can largely be seen as a process of gradually generalizing advanced algorithmic techniques from ``simple'' regions with strong structure and loose constraints to ``hard'' regions with greater universality on this ``difficulty spectrum.''

\subsection{Fundamental Hardness and Inapproximability}

The computational hardness of CVRP is not limited to being NP-hard but is also reflected in the difficulty of its approximation, which sets theoretical boundaries for research on its approximation algorithms.

\textbf{NP-Hardness:} As previously mentioned, when the number of customers a vehicle can serve is $Q \ge 3$, CVRP is NP-hard \cite{hassin2005complexity, zhao2022improved}. More importantly, due to its connection with the Bin Packing Problem, CVRP is APX-hard. This means that even in the simplest one-dimensional Euclidean space, no Polynomial Time Approximation Scheme (PTAS) exists unless P=NP. A PTAS is a class of algorithms that, for any given $\epsilon > 0$, can find a solution with a cost no more than $(1+\epsilon)$ times the optimal solution within polynomial time.

\textbf{Inapproximability in General Metric Spaces:} For CVRP in general metric spaces, a stronger inapproximability result exists. This result stems from its parent problem—Metric TSP. In a landmark work in 1992, Arora et al. \cite{arora1992proof} proved that for any $\epsilon > 0$, approximating Metric TSP within a factor of $(1 + \frac{1}{129} - \epsilon)$ in polynomial time is NP-hard. Since any CVRP instance, if its capacity is large enough, is equivalent to a TSP instance, this inapproximability of TSP is directly transferred to CVRP. This conclusion is a significant watershed in CVRP approximation algorithm research: it explicitly points out that seeking approximate solutions of arbitrary precision in general metric spaces is futile, thus directing the focus of research to two main directions: one is to seek the best constant-factor approximation algorithms in general metric spaces; the other is to explore the possibility of obtaining PTAS or QPTAS in metric spaces with special structures (such as geometric structures).

\section{The Foundational Iterated Tour Partitioning (ITP) Framework}

In the grand landscape of CVRP approximation algorithms, the Iterated Tour Partitioning (ITP) framework \cite{haimovich1985bounds} occupies a central and enduring position. Proposed by Haimovich and Rinnooy Kan in 1985, ITP is a classic embodiment of the ``route-first, cluster-second'' strategy. Its algorithmic idea is concise and robust; for over thirty years, it has been the cornerstone of the best-performing approximation algorithms for solving CVRP in general metric spaces. Recent research breakthroughs have not discarded ITP but rather optimized its input through deeper structural insights to enhance overall performance.

\subsection{Core Component: Christofides-Serdyukov TSP Algorithm}

The first step of the ITP framework is to solve a TSP instance; therefore, a high-quality TSP approximation algorithm is the guarantee of the entire framework's performance. In metric spaces, the algorithm proposed independently by Christofides (1976) \cite{christofides1976worst, christofides2022worst} and Serdyukov (1976) has long been the gold standard. For a historical perspective on this algorithm, see \cite{vanbevern2020historical}. This algorithm can provide a solution with an approximation ratio of $3/2$ for Metric TSP in polynomial time, a record that has stood for nearly half a century. Its algorithmic steps are concise:

\begin{enumerate}
    \item Compute a Minimum Spanning Tree (MST) on the graph formed by all customers and the depot.
    \item Identify all vertices with odd degrees in the MST. According to the handshaking lemma, there must be an even number of such vertices.
    \item Compute a minimum-weight perfect matching on these odd-degree vertices.
    \item Merge the edges of the MST and the perfect matching to form a multigraph where all vertices have even degrees.
    \item Find an Eulerian tour in this multigraph.
    \item Eliminate repeated visits to vertices by performing ``shortcuts'' in the Eulerian tour to obtain a Hamiltonian cycle, i.e., the TSP solution.
\end{enumerate}

We denote the performance ratio of the TSP approximation algorithm as $\rho$. For the Christofides-Serdyukov algorithm, $\rho = 3/2$. Although Karlin et al. \cite{karlin2021slightly} achieved a minor improvement on this ratio in 2021, $\rho = 3/2$ is still used as the benchmark in most CVRP analyses.

\subsection{Technical Details of ITP Algorithm}

The flow of the ITP algorithm clearly embodies the idea of decomposing the complex CVRP into two more manageable sub-problems: an unconstrained ordering problem (TSP) and a one-dimensional partitioning problem.

\textbf{Algorithm Steps:}

\begin{enumerate}
    \item \textbf{Construct ``Giant Tour'':} Treat all customers as a whole and use a TSP algorithm with an approximation ratio of $\rho$ (such as the Christofides-Serdyukov algorithm) to calculate a ``giant'' Hamiltonian cycle passing through all customers. In this step, the vehicle capacity limit is completely ignored.
    \item \textbf{Optimal Partitioning:} Partition the giant tour (a sequence of customers) obtained in the previous step. The goal is to cut it into the minimum number of continuous sub-segments while ensuring that the total demand of all customers in each sub-segment does not exceed the vehicle capacity $Q$. This one-dimensional partitioning problem is a classic dynamic programming problem that can be solved optimally in polynomial time.
    \item \textbf{Form Final Routes:} For each customer sub-segment obtained by partitioning, connect its first and last customer nodes to the depot to form a complete vehicle route satisfying capacity constraints. The collection of all these routes constitutes a feasible solution to the CVRP problem.
\end{enumerate}

\subsection{Classical Worst-Case Analysis}

The elegance of the ITP algorithm lies in its rigorous theoretical performance guarantees. Its approximation ratio is a function of the TSP algorithm's approximation ratio $\rho$ and the vehicle capacity. The classical analysis results are as follows:

\begin{itemize}
    \item For CVRP with \textbf{Splittable Demand} and \textbf{Unit Demand}, the approximation ratio of ITP is $\rho + 1 - \frac{1}{Q}$. Intuitively, the cost increased by the partitioning process mainly comes from connecting the endpoints of each sub-segment back to the depot, and this part of the cost is roughly equivalent to an additional optimal TSP path.
    \item For the more challenging CVRP with \textbf{Unsplittable Demand}, the approximation ratio is $\rho + 1 - \frac{1}{Q}$. The additional cost factor comes from potentially worse partitioning schemes generated when handling unsplittable demands.
\end{itemize}

Haimovich and Rinnooy Kan's original analysis provided a more refined bound. Let $\bar{d}$ denote the average demand. When the number of customers $n$ is a multiple of $Q$ (for unit demand), the approximation ratio approaches $\alpha = 1 + (1 - 1/Q)\rho$. When using the Christofides-Serdyukov algorithm ($\rho=1.5$), this formula leads to a bound around $2.5$, which aligns with classical results.

\subsection{Modern Improvements: Structural Analysis by Blauth, Traub, and Vygen}

For over thirty years, the aforementioned classical analysis based on ITP represented the highest level of approximation algorithms for CVRP in general metric spaces. It was not until 2023 that Blauth, Traub, and Vygen \cite{blauth2023improving} achieved a fundamental breakthrough. They did not propose a completely new algorithm but improved its upper bound performance through a profound rethinking of the ITP framework.

Their work centers on a refined structural characterization of the ``difficult instances''—the instances where the ITP algorithm performs worst. They discovered that when the ITP algorithm's solution is far inferior to the optimal solution, the paths of the optimal solution must exhibit a specific geometric structure.

\textbf{Core Insight:} In these difficult instances, almost every optimal path possesses two key features:
\begin{enumerate}
    \item \textbf{Small Detour:} The length of each path differs very little from the round-trip distance from the depot to its ``peak'' (the customer on the path furthest from the depot). This implies that the two sub-paths from the depot to the peak constituting the path are essentially shortest paths, with no significant ``detour.''
    \item \textbf{Clustered Demand:} The total demand of each path is almost equal to the vehicle capacity $Q$, and the vast majority of the demand is concentrated near the peak point.
\end{enumerate}

This insight reveals the fundamental weakness of the classic ITP method: an optimal TSP ``giant tour'' is not necessarily a suitable skeleton to be partitioned into CVRP paths satisfying the above ``small detour'' and ``clustered demand'' characteristics. The goal of the optimal TSP tour is global shortness, which might disrupt the excellent structure of local paths to connect a distant customer.

\textbf{Improved Method:} The new algorithm designed by Blauth et al. \cite{blauth2023improving} is precisely aimed at exploiting this structural weakness. Their algorithm can be seen as constructing a more ``friendly'' input tour for the ITP framework:
\begin{enumerate}
    \item \textbf{Identify and Guess:} The algorithm first attempts to identify potential ``clusters'' among customers and ``guesses'' the ``target groups'' where the peaks of the optimal paths might be located.
    \item \textbf{Path Construction and Matching:} Next, instead of constructing a single giant tour, the algorithm computes a set of paths starting from the depot and ending at these target groups. Through ingenious mathematical programming or combinatorial algorithms, it ensures that the total length and total detour of these paths are as small as possible.
    \item \textbf{Generate New Tour:} Finally, by computing a low-cost matching between the endpoints of these paths, they are connected into one or more Eulerian tours, thereby forming a new, structurally superior giant tour.
    \item \textbf{Final Partitioning:} Input this newly generated, more ``partition-friendly'' giant tour into the classic ITP partitioning procedure.
\end{enumerate}

In this way, they successfully improved the approximation ratio for unsplittable CVRP from $\alpha \approx 3.5$ to a significantly lower value, where $\alpha$ is a small constant greater than 1. This milestone work demonstrates that the key to improving CVRP approximation performance may lie not in completely abandoning ITP, but in how to more intelligently construct an initial ordering that preserves the intrinsic structure of the optimal CVRP solution.

\section{Approximation Schemes in Geometric Spaces}

When the context of CVRP shifts from general metric spaces to Euclidean spaces with geometric structures, the toolbox for algorithm design is greatly enriched. Geometric properties make powerful techniques like recursive space decomposition possible, thereby opening the door to designing approximation schemes (PTAS or QPTAS) with approximation ratios arbitrarily close to 1 \cite{bern1997approximation}. The history of this field clearly demonstrates how a breakthrough idea for a parent problem (TSP) is gradually adapted and deepened to cope with the additional complexity brought by the sub-problem (CVRP).

\subsection{Theoretical Cornerstone: Arora's Euclidean TSP Framework}

In 1998, the PTAS proposed by Sanjeev Arora \cite{arora1998polynomial} for Euclidean TSP stood as a monument in the field of geometric approximation algorithms and laid the methodological foundation for subsequent CVRP research. The core idea of Arora's algorithm is to utilize the separability of geometric space:

\begin{enumerate}
    \item \textbf{Recursive Space Decomposition:} The algorithm performs a recursive, randomized quadtree (or k-d tree) decomposition on the bounding box containing all customer points until the number of customer points within each minimum cell becomes very small.
    \item \textbf{Portals:} On the boundary of each cell formed by the decomposition, the algorithm places a finite number of ``portal'' points. Any path crossing the cell boundary is required to pass through these portal points.
    \item \textbf{Dynamic Programming:} Based on this portal structure, the algorithm uses dynamic programming to compute the optimal (or approximately optimal) path passing through these cells and portals. By limiting the number of portals (where the number depends only on $\epsilon$ and not the input size $n$), the state space of dynamic programming is controlled, thereby ensuring the algorithm's overall running time is polynomial.
\end{enumerate}

In this way, for any given $\epsilon$, Arora's algorithm can find a TSP path with a cost no more than $(1+\epsilon)$ times the optimal solution in polynomial time.

\subsection{Early Applications to CVRP}

Arora's framework was soon attempted for solving Euclidean CVRP, but initial success was limited to scenarios with either very weak or very strong capacity constraints.

\begin{itemize}
    \item When the vehicle capacity $Q$ is very large, e.g., $Q \ge n$, an optimal CVRP solution likely contains only one path. In this case, the problem is almost identical to TSP, so Arora's PTAS can be directly applied, followed by the ITP framework (where the partitioning step is trivial) to obtain a $(1+\epsilon)$-approximate solution.
    \item For cases where capacity $Q$ is small, Asano, Katoh, and Kawashima successfully extended Arora's technique in a 1997 work \cite{asano1997covering} to design a PTAS for CVRP with $Q = O(\log n)$ (where $Q$ is constant). This indicates that when capacity grows slowly, Arora's dynamic programming framework can still be adapted.
\end{itemize}

However, for arbitrary capacity $Q$, effectively generalizing Arora's framework remained an unresolved difficulty. The fundamental challenge lies in the fact that CVRP paths can arbitrarily start and end at the depot, which differs significantly from the single Hamiltonian cycle structure of TSP, causing the intersection of paths with space decomposition boundaries to become exceptionally complex, and the dynamic programming states to explode accordingly.

\subsection{Milestone Progress: QPTAS by Das and Mathieu}

In 2010, Das and Mathieu \cite{das2010quasi} achieved a major breakthrough by proposing the first Quasi-Polynomial Time Approximation Scheme (QPTAS) for Euclidean CVRP with arbitrary capacity. The running time of QPTAS is of the form $n^{O(\log n)}$, which, although not strictly polynomial time, is very close.

Das and Mathieu's algorithm inherited Arora's recursive space decomposition idea, but its core innovation lay in introducing a new method to handle the complex interaction between capacity constraints and space decomposition:

\begin{enumerate}
    \item \textbf{Structure Theorem and Guessing:} They first proved a structural theorem regarding optimal CVRP solutions, showing that any optimal solution can be modified into a ``well-structured'' solution with little increase in cost. In this well-structured solution, the number of path segments crossing decomposition cell boundaries is limited.
    \item \textbf{``Paying'' for Complexity:} The core idea of the algorithm is to ``guess'' the ``profile'' of paths as they cross cell boundaries, such as the accumulated demand on the path. To keep the number of guesses within a quasi-polynomial range, they imposed extra cost penalties in the dynamic programming for those path segments with ``rare'' or ``complex'' profiles. This amounts to trading a portion of solution quality (allowing a slight increase in cost) for a drastic compression of the dynamic programming state space.
\end{enumerate}

Through this ingenious mechanism of ``paying for complexity,'' Das and Mathieu successfully controlled the number of dynamic programming states at a quasi-polynomial level, thereby constructing the first approximation scheme applicable to arbitrary capacities.

Building upon this foundational work, Khachay, Ogorodnikov, and Khachay \cite{khachay2021efficient} generalized the approximation scheme to operate in metric spaces of any fixed doubling dimension ($d>1$). While a straightforward application of Das and Mathieu’s original exhaustive search step over interface vectors would lose its quasi-polynomial efficiency in these broader metric spaces, Khachay et al. resolved this by introducing a novel internal dynamic programming algorithm to replace the exhaustive search. Through this crucial refinement, they successfully retained a QPTAS for these geometric settings, provided that the vehicle capacity does not exceed $\text{polylog}(n)$.

\subsection{The Journey Toward Full PTAS}

Das and Mathieu's QPTAS opened a new path for the field, and subsequent research efforts have been dedicated to eliminating the quasi-polynomial time bottleneck based on their framework through more refined analysis, thereby realizing a true PTAS under broader conditions.

\begin{itemize}
    \item \textbf{Adamaszek, Czumaj, and Lingas (2010):} Building on the Das-Mathieu framework, they designed the first PTAS for two-dimensional Euclidean CVRP \cite{adamaszek2010ptas}, but its scope was limited to ``moderately large'' capacities, i.e., when vehicle capacity $Q \le 2^{\log^{\delta} n}$, where $\delta$ is a small constant dependent on $\epsilon$. Their technical key was proving that under this capacity limit, the profiles of paths crossing boundaries could be more efficiently classified and managed, reducing the running time to a polynomial level.
    \item \textbf{Khachay and Zaytseva (2015):} Generalized Adamaszek et al.'s results to three-dimensional Euclidean space \cite{khachay2015polynomial}, also achieving a PTAS under the capacity limit of $Q \le 2^{\log^{\delta} n}$.
    \item \textbf{Khachay and Ogorodnikov (2019 and later):} Khachay's team continued to optimize and expand this framework in a series of works, achieving significant results particularly on the more complex CVRPTW variant. They designed an Efficient Polynomial Time Approximation Scheme (EPTAS) for Euclidean CVRPTW \cite{khachai2019polynomial, khachay2019polynomial2}, including cases with non-uniform demands \cite{khachay2019approximation}, with a running time of $O(n \log n)$, but requiring the number of time windows $M$ and capacity $Q$ to be constants. When $M$ and $Q$ are not constants but satisfy $Q \cdot M = O(\log n)$, their algorithm can also achieve PTAS performance.
\end{itemize}

The evolutionary path of this series of research clearly shows that from PTAS for TSP to QPTAS for CVRP, and then to conditional PTAS, the core intellectual challenge lies in how to quantify and control the combinatorial complexity introduced by the extra dimension of capacity constraints. Every step of progress stems from a deeper understanding of the structure of optimal solutions under geometric decomposition.

\section{Algorithms on Structured and General Graphs}

Beyond Euclidean space, research on CVRP approximation algorithms has unfolded along two main technical paths. One path focuses on exploiting the special combinatorial structures of graphs, such as trees and planar graphs, to obtain better performance guarantees through specially designed algorithms or metric embedding techniques. The other path returns to general metric spaces, using powerful general optimization tools like linear programming relaxation to challenge and improve the classic ITP framework. These two paths represent two different philosophies in current approximation algorithm research: the former tames complexity by restricting the scope of problem instances, while the latter attempts to directly attack the problem in its most general form with stronger mathematical weapons.

\subsection{CVRP on Tree Metrics}

Trees are the simplest metric space aside from a straight line; their unique structure (unique path between any two points) greatly simplifies the path planning problem, making them an ideal testing ground for new algorithmic ideas.

\begin{itemize}
    \item \textbf{Constant Factor Approximation:} For CVRP on trees, early research achieved approximation ratios better than those for general metric spaces. Asano et al. provided a $1.5$-approximation algorithm for Splittable CVRP on trees in 2001 \cite{asano2001new}. Subsequently, Becker improved this result in 2018 \cite{becker2018tight} to a tight $4/3$-approximation algorithm, reaching the theoretical limit known for this problem based on specific lower bounds (see also the framework in \cite{becker2019framework}).
    \item \textbf{Implementation of PTAS:} In recent years, Grandoni, Mathieu, and Zhou have made a series of decisive advances in research on CVRP on trees. They first designed a $(1.5+\epsilon)$-approximation algorithm for CVRP with unsplittable demands \cite{grandoni2022unsplittable} (later tightened in \cite{mathieu2022tight}) and eventually developed a PTAS for CVRP with unit demands and splittable demands on trees \cite{mathieu2023ptas}. This series of results indicates that in such a highly structured metric space as a tree, the intrinsic complexity of CVRP is significantly weakened, allowing for solutions approximating arbitrary precision.
\end{itemize}

\subsection{Metric Embedding Techniques}

Metric embedding is a powerful modern algorithm design paradigm. Its core idea is to embed a ``complex'' metric space $M$, which is difficult to handle directly, into a structurally ``simpler'' host space $M'$ (easier for algorithm design) via a mapping $f$, while ensuring that the original distance information is not lost significantly.

\begin{itemize}
    \item \textbf{Core Concept:} The quality of the mapping $f$ is measured by its ``distortion,'' i.e., the maximum degree to which it stretches or compresses the distance between any two points. If a low-distortion embedding into a simple space (like a tree or a graph with bounded treewidth) can be found in polynomial time, then efficient algorithms (like dynamic programming) can be used to solve the problem in the host space, and the solution can be mapped back to the original space to obtain an approximate solution to the original problem.
    \item \textbf{PTAS for Bounded Highway Dimension:} Becker, Klein, and Saulpic \cite{becker2017polynomial} utilized metric embedding techniques in 2017 to design the first PTAS for CVRP on a class of graphs known as ``Bounded Highway Dimension graphs.'' Highway dimension is a metric measuring the ``tree-likeness'' of a graph metric; many real-world road networks have small highway dimensions. Their key technical contribution was a novel metric embedding method capable of embedding a bounded highway dimension graph into a graph with bounded treewidth (extended to other structures like bounded doubling dimension in \cite{jayaprakash2023approximation, khachay2020efficient, ogorodnikov2021approximation}). Specifically, the error of this embedding is \textit{additive} $\epsilon D$, where its magnitude is proportional to the sum of the distances of the two points to the depot, i.e., $\epsilon (d(u, 0) + d(v, 0))$. Since CVRP can be efficiently solved by dynamic programming on bounded treewidth graphs, this high-quality embedding directly led to a PTAS, provided that the vehicle capacity $Q$ is bounded.
    \item \textbf{Technical Extension:} This powerful embedding idea was subsequently further generalized. Becker, Klein, and Schild \cite{becker2019ptas} applied it to planar graphs in 2019, providing a PTAS for Planar CVRP with bounded capacity. Recently, Vincent Cohen-Addad et al. (2023) \cite{cohen2023planar} extended this technique to a broader class of minor-free graphs, utilizing tools such as light spanners \cite{cohen2020light}, providing a QPTAS for CVRP on such graphs with arbitrary capacity. Recent work has also explored graphic metrics \cite{momke2023capacitated}.
\end{itemize}

\subsection{Linear Programming Relaxation and Rounding}

Linear Programming (LP) relaxation and rounding is a classic and powerful general technique in approximation algorithm design. Its basic flow is:

\begin{enumerate}
    \item Formulate the combinatorial optimization problem as an Integer Linear Program (ILP).
    \item ``Relax'' the integer constraints in the ILP (e.g., relax variable $x \in \{0, 1\}$ to $0 \le x \le 1$) to obtain an LP that can be solved in polynomial time.
    \item Solve the LP to obtain an optimal solution that may contain fractional values.
    \item Design a ``rounding'' algorithm to convert this fractional solution into an integer solution that satisfies the original problem constraints.
\end{enumerate}

The key and difficulty of the entire method lie in designing a rounding strategy such that the cost increase of the final integer solution relative to the LP optimal value (which is a lower bound for the optimal solution of the original problem) is as small as possible. For CVRP, common ILP models include the two-index flow formulation and the set partitioning formulation.

\textbf{Application in CVRP:} Friggstad et al. \cite{friggstad2021improved} successfully applied the LP rounding method in 2021 to design an algorithm with an approximation ratio of approximately $3.19$ (specifically $\ln(1.5) + 1 \approx 1.41$ plus terms related to TSP) for unsplittable CVRP in general metric spaces. This result provides another distinct path for improving the upper bound performance of the classic ITP framework. It does not rely on combinatorial structural analysis of ``difficult instances'' but appeals to the powerful capabilities of mathematical programming. This indicates that even in the most general metric space setting, there is still room to improve the approximation ratio through stronger LP relaxations and more ingenious randomized rounding techniques.

\section{Conclusion: Current State of the Art and Future Directions}

After decades of development, the field of CVRP approximation algorithms has formed a rich and profound theoretical system. From the foundational ITP framework to ingenious approximation schemes for geometric spaces, and then to the use of modern tools like metric embedding and LP relaxation to handle structured and general graphs, researchers have drawn a detailed ``approximability map'' for this classic NP-hard problem. This section will summarize the current technological level and highlight major open problems that continue to challenge the wisdom of the field.

\subsection{Overview of Current Best Approximation Results}

To clearly demonstrate the highest achievements of current CVRP approximation algorithms, the following table (Table 2) systematically summarizes the optimal approximation ratios under different metric spaces and demand type settings. This table not only lists performance guarantees but also indicates the core technical paradigms upon which these guarantees rely, thereby highlighting the developmental lineage of the field elucidated in this survey.

\begin{longtable}{p{0.2\textwidth} p{0.15\textwidth} p{0.15\textwidth} p{0.15\textwidth} p{0.15\textwidth} p{0.15\textwidth}}
\caption{Summary of Current Best Results for CVRP Approximation Algorithms} \label{tab:summary} \\
\toprule
\textbf{Metric Space} & \textbf{Demand Type} & \textbf{Capacity ($Q$ or $k$)} & \textbf{Best Approx. Ratio} & \textbf{Core Technology} & \textbf{Key Reference} \\
\midrule
\endfirsthead
\caption[]{Summary of Current Best Results (continued)} \\
\toprule
\textbf{Metric Space} & \textbf{Demand Type} & \textbf{Capacity ($Q$ or $k$)} & \textbf{Best Approx. Ratio} & \textbf{Core Technology} & \textbf{Key Reference} \\
\midrule
\endhead
\bottomrule
\endfoot
\bottomrule
\endlastfoot
\textbf{General Metric} & Unit / Splittable & Arbitrary & $\alpha \approx 2.5$ & Improved ITP (Structural Analysis) & Blauth, Traub, \& Vygen (2023) \cite{blauth2023improving} \\
\textbf{General Metric} & Unsplittable & Arbitrary & $\approx 3.19$ & LP Relaxation \& Rounding & Friggstad et al. (2021) \cite{friggstad2021improved} \\
\textbf{General Metric} & Unit / Splittable & Arbitrary & $\approx 2.5$ & Improved ITP & Blauth, Traub, \& Vygen (2023) \cite{blauth2023improving} \\
\textbf{Euclidean Space ($\mathbb{R}^2$)} & Unsplittable & Arbitrary & QPTAS & Geometric Recursive Decomposition & Das \& Mathieu (2010) \cite{das2010quasi} \\
\textbf{Euclidean Space ($\mathbb{R}^2$)} & Unsplittable & $Q \le 2^{\log^\delta n}$ & PTAS & Geometric Recursive Decomposition & Adamaszek et al. (2010) \cite{adamaszek2010ptas} \\
\textbf{Tree Metric} & Unit / Splittable & Arbitrary & PTAS & Dynamic Programming & Mathieu \& Zhou (2023) \cite{mathieu2023ptas} \\
\textbf{Tree Metric} & Unsplittable & Arbitrary & $1+\epsilon$ & Combinatorial Algorithms & Mathieu \& Zhou (2022) \cite{grandoni2022unsplittable} \\
\textbf{Bounded Highway Dim.} & Unsplittable & Bounded $Q$ & PTAS & Metric Embedding & Becker, Klein, \& Saulpic (2017) \cite{becker2017polynomial} \\
\textbf{Planar Graphs} & Unsplittable & Bounded $Q$ & PTAS & Metric Embedding & Becker, Klein, \& Schild (2019) \cite{becker2019ptas} \\
\textbf{Minor-Free Graphs} & Unsplittable & Arbitrary & QPTAS & Metric Embedding & Cohen-Addad et al. (2023) \cite{cohen2023planar} \\
\end{longtable}

\textit{Note:} $\rho$ represents the approximation ratio of the TSP subroutine used, typically taking $\rho=1.5$ from the Christofides-Serdyukov algorithm. PTAS stands for Polynomial Time Approximation Scheme, and QPTAS stands for Quasi-Polynomial Time Approximation Scheme.

\subsection{Core Open Problems and Future Directions}

Despite immense progress, several fundamental and highly challenging problems remain unresolved in the field of CVRP approximation algorithms. These problems not only possess profound theoretical value but also guide future research directions.

\begin{enumerate}
    \item \textbf{Does a PTAS exist for Euclidean CVRP?} This is undoubtedly the most central and compelling open problem in the field. Das and Mathieu's QPTAS \cite{das2010quasi} shows that for Euclidean CVRP with arbitrary capacity, we can obtain approximation ratios arbitrarily close to 1, but at the cost of quasi-polynomial running time. Removing this ``quasi-polynomial'' barrier, i.e., designing a PTAS with a running time of $n^{O(1)}$, is the ``Holy Grail'' of this field. Solving this problem may require fundamental innovations to the recursive decomposition frameworks of Arora and Das-Mathieu, or the introduction of entirely new geometric algorithmic ideas.
    \item \textbf{Narrowing the approximation ratio gap for CVRP in General Metric Spaces.} For unsplittable CVRP in general metric spaces, a huge gap still exists between the current best approximation ratio (slightly below 3.5) and the known APX-hardness lower bound (1.5). The combinatorial method of Blauth et al. \cite{blauth2023improving} and the LP rounding method of Friggstad et al. \cite{friggstad2021improved} represent two different paths of attack. Future research may continue to deepen along these two directions: Can the input to ITP be further optimized through more refined structural decomposition of ``difficult instances''? Or, can stronger LP relaxations (e.g., based on column generation or more complex cutting planes) and more effective rounding schemes be designed? Combining these two technical ideas could also be a fruitful direction.
    \item \textbf{Designing PTAS for Planar and Minor-Free Graphs.} Currently, for CVRP on planar and minor-free graphs, we only possess PTAS when vehicle capacity is bounded, while only QPTAS is available for arbitrary capacity. Considering the ubiquity of such graphs in the real world, designing a PTAS applicable to arbitrary capacity for them has significant theoretical and practical implications. This may require developing more powerful metric embedding techniques capable of handling arbitrary capacity demands, or combining geometric decomposition ideas more closely with the combinatorial properties of graphs (such as separator theorems).
    \item \textbf{Developing strong approximation algorithms for key CVRP variants.} Compared to classic CVRP, research on approximation algorithms for many important real-world variants, such as CVRP with Time Windows (CVRPTW), is far from mature. Although Khachay et al. have achieved preliminary PTAS/EPTAS results for CVRPTW in Euclidean space, we know very little about the approximability of CVRPTW in more general metric spaces. Additionally, for variants closer to real-world scenarios like CVRP with Stochastic Demands (Stochastic CVRP) \cite{mathieu2021probabilistic, nie2023euclidean}, or Location Routing \cite{heine2023bifactor}, designing approximation algorithms with rigorous performance guarantees is an emerging research frontier full of opportunities and challenges.
\end{enumerate}

In summary, research on CVRP approximation algorithms is in an exciting phase. On one hand, core difficulties of classic problems (like Euclidean CVRP) remain to be conquered; on the other hand, new algorithmic tools (like metric embedding) and new problem perspectives (like structured graphs and stochastic variants) are constantly expanding the boundaries of research in this field. Future breakthroughs will likely come from the intersection and fusion of multiple fields such as geometry, graph theory, combinatorial optimization, and mathematical programming.

\smallskip

\medskip

\vskip 0.2cm
\noindent\rule{\textwidth}{0.5pt}
\vskip  0.2cm

\medskip


\smallskip

\medskip

{\bf INFORMATION ABOUT THE AUTHORS}
\smallskip

\noindent{\bf Yongyu Chen} -- Department of Mathematics and Computer Science, Ural Federal University, Yekaterinburg, Russia


\end{document}